\documentclass[epj,final]{svjour}
\usepackage{graphics}
\begin{document}
\title{Self-consistent thermodynamics for the Tsallis statistics in the grand canonical ensemble: Nonrelativistic hadron gas}
\author{A.S.~Parvan 
}                     
%
%
\institute{Bogoliubov Laboratory of Theoretical Physics, Joint Institute for Nuclear Research, 141980 Dubna, Russian Federation \and
Department of Theoretical Physics, Horia Hulubei National Institute of Physics and Nuclear Engineering, 077125 Bucharest, Romania \and
Institute of Applied Physics, Moldova Academy of Sciences, MD-2028 Chisinau, Republic of Moldova}
\date{Received: date / Revised version: date}
%
\abstract{In the present paper, the Tsallis statistics in the grand canonical ensemble was reconsidered in a general form. The thermodynamic properties of the nonrelativistic ideal gas of hadrons in the grand canonical ensemble was studied numerically and analytically in a finite volume and the thermodynamic limit. It was proved that the Tsallis statistics in the grand canonical ensemble satisfies the requirements of the equilibrium thermodynamics in the thermodynamic limit if the thermodynamic potential is a homogeneous function of the first order with respect to the extensive variables of state of the system and the entropic variable $z=1/(q-1)$ is an extensive variable of state. The equivalence of canonical, microcanonical and grand canonical ensembles for the nonrelativistic ideal gas of hadrons was demonstrated.
} 
\titlerunning{Self-consistent thermodynamics for the Tsallis statistics in the grand canonical ensemble}
\authorrunning{A.S.~Parvan}
\maketitle
\section{Introduction}\label{intro}
The Tsallis power-law distribution describes very well the experimental data in both high-energy physics~\cite{Cleymans1,Biro0,Biro1} and astrophysics~\cite{Esquivel1,Betzler1}. In particular, the Tsallis-like distributions are successfully used to describe the transverse momentum spectra of the particles produced in the proton-proton and nucleus-nucleus collisions at the LHC and RHIC energies~\cite{Alice1,Alice1a,Alice1b,Atlas1,Cms3,Rybczynski14,Cleymans13,Parvan14}. However, until now the theoretical foundation of the Tsallis statistics~\cite{Tsal88,Tsal98} is far from a definitive solution. The Tsallis generalized statistical mechanics, by definition, is in the thermal equilibrium. Thus, the proof of the thermodynamic self-consistency of the Tsallis statistics remains a primordial task, although some progress has been made. In this direction, there exist some approaches~\cite{Abe0,Wang2,Abe1,Parv1,Parv2,Parv2a}.

The main aim of this paper is to prove analytically and numerically the thermodynamic self-consistency of the Tsallis statistics in the grand canonical ensemble in the approach when the entropic variable $z=1/(q-1)$ is considered to be an extensive variable of state and to apply this approach of the Tsallis statistics to the hadron ideal gas. We investigate the grand canonical ensemble because in the high energy collisions the number of particles are not exactly conserved and, therefore, the thermodynamic quantities and the transverse momentum distributions of hadrons in high-energy processes are usually described by this statistical ensemble.

The statistical mechanics defined in a thermal equilibrium is thermodynamically self-consistent, i.e., it agrees with the requirements of the equilibrium thermodynamics, if the thermodynamic potential of the statistical ensemble is a homogeneous function of the first order with respect to the extensive variables of state of the system in the thermodynamic limit~\cite{Prigogine,Kvasnikov,Parv2,Parv2a}. The thermodynamic self-consistence of the Tsallis statistics in the canonical and microcanonical ensembles was proved in~\cite{Parv2,Parv2a}. It was shown that the homogeneous property of the thermodynamic potential provides the zeroth law of thermodynamics, the principle of additivity, the Euler theorem and the Gibbs-Duhem relation if the entropic variable $z$ is an extensive variable of state. Note that the scaling property of variable $z$ in the Tsallis statistics and its relation to the thermodynamic limit was firstly discussed in~\cite{Botet2,Botet1}.

At present, the concept of the relativistic hadron ideal gas is extensively applied in many approaches of high-energy physics. It is used to study the hadron production mechanism and the QCD phase transition in the heavy-ion collisions at high energies~\cite{Andronic1,Becattini1,Braun1,Andronic2}. However, in the present paper we study the thermodynamic properties of the hadron gas in the Tsallis statistics in the simple case of the hadron gas in the nonrelativistic limit.

The structure of the paper is as follows. In Section~\ref{sec:1}, we formulate the Tsallis statistics in the grand canonical ensemble. In Sections~\ref{sec:2} and \ref{sec:3}, we consider the nonrelativistic ideal gas in a finite volume and the themodynamic limt, respectively. The ensemble equivalence is demonstrated in Section~\ref{sec:4}. The numerical calculations for the hadron ideal gas are given in Section~\ref{sec:5}, and the main conclusions are summarized in the final section.

\section{Tsallis statistics in the grand canonical ensemble}\label{sec:1}
Let us consider the grand canonical ensemble of the dynamical systems at the constant temperature $T$, the volume $V$, the chemical potential $\mu$ and the thermodynamic coordinate $z$ in a thermal contact with a heat bath. The system exchanges with its surroundings the energy and matter. The entropy of the Tsallis statistics in the grand canonical ensemble is~\cite{Tsal88,Tsal98}
\begin{eqnarray}\label{1}
    S &=&  \sum\limits_{i} \delta_{V,V_{i}}  p_{i} S_{i}, \\ \label{2}
    S_{i} &=& k_{B} z (1-p_{i}^{1/z}),
\end{eqnarray}
where $p_{i}$ is the probability of the $i$-th microstate of the system, $k_{B}$ is the Boltzmann constant, $z=1/(q-1)$ and $q\in\mathbf{R}$ is a real parameter taking values $0<q<\infty$. In the limit $|z|\to\infty$ or $q\to 1$, the entropy (\ref{1}) recovers the Boltzmann-Gibbs entropy, $S=-k_{B} \sum_{i} p_{i} \ln p_{i}$. Note that throughout the paper we use the system of natural units $\hbar=c=k_{B}=1$.

The thermodynamic potential of the grand canonical ensemble is related to the fundamental thermodynamic potential (energy) by the Legendre transform~\cite{Arnold} and can be written as
\begin{eqnarray}\label{3}
  \Omega &=& \langle H\rangle - TS - \mu \langle N\rangle = \sum\limits_{i} \delta_{V,V_{i}}  p_{i} \Omega_{i}, \\ \label{4}
  \Omega_{i} &=& E_{i}-\mu N_{i} -T z (1-p_{i}^{1/z}),
\end{eqnarray}
where $E_{i}$ and $N_{i}$ are the energy and number of particles, respectively, in the $i$-th microscopic state of the system. The probabilities of microstates in the grand canonical ensemble are constrained by an additional function
\begin{equation}\label{5}
    \varphi=\sum\limits_{i} \delta_{V,V_{i}} p_{i} - 1 = 0.
\end{equation}

The unknown probabilities $\{p_{i}\}$ are obtained from the constrained local extrema of the thermodynamic potential (\ref{3}) by the method of the Lagrange multipliers~\cite{Krasnov}
\begin{eqnarray}\label{6}
 \Phi &=& \Omega - \lambda \varphi,  \\ \label{7}
  \frac{\partial \Phi}{\partial p_{i}} &=& 0,
\end{eqnarray}
where $\lambda$ is an arbitrary real constant. Substituting Eqs.~(\ref{3}) and (\ref{5}) into Eqs.~(\ref{6}), (\ref{7}) we obtain
\begin{equation}\label{8}
  \Omega_{i}+p_{i}\frac{\partial \Omega_{i}}{\partial p_{i}} -\lambda=0.
\end{equation}
Substituting Eq.~(\ref{4}) into (\ref{8}) and using Eq.~(\ref{5}), we obtain the normalized probabilities of the grand canonical ensemble as
\begin{eqnarray}\label{9}
p_{i} &=& \left[1+\frac{1}{z+1}\frac{\Lambda-E_{i}+\mu N_{i}}{T}\right]^{z}, \\ \label{10}
    1 &=& \sum\limits_{i} \delta_{V,V_{i}} \left[1+\frac{1}{z+1}\frac{\Lambda-E_{i}+\mu N_{i}}{T}\right]^{z},
\end{eqnarray}
where $\Lambda\equiv \lambda-T$ and $\partial E_{i}/\partial p_{i}=0$. In the limit $|z|\to\infty$, the probability $p_{i}=\exp[(\Lambda-E_{i}+\mu N_{i})/T]$ and $\Lambda=-T\ln Z$, where $Z=\sum_{i} \exp[-(E_{i}-\mu N_{i})/T]$.

Substituting the probabilities (\ref{9}) into Eqs.~(\ref{3}), (\ref{4}) and using Eq.~(\ref{10}) we obtain the thermodynamic potential of the grand canonical ensemble as
\begin{equation}\label{11}
    \Omega = \Lambda + T \left[1-\sum\limits_{i} \delta_{V,V_{i}} p_{i}^{1+\frac{1}{z}}\right].
\end{equation}
Taking the first derivative of the function $\Omega$ with respect to temperature $T$ and using Eqs.~(\ref{9}), (\ref{10}), and the relations $\partial E_{i}/\partial T=0$, $\partial N_{i}/\partial T=0$ we obtain the entropy (\ref{1})--(\ref{2}) of the system as
\begin{equation}\label{12}
   S =-\frac{\partial \Omega}{\partial T} = \sum\limits_{i} \delta_{V,V_{i}}  p_{i} S_{i}.
\end{equation}
Applying the Legendre back-transformation to the function $\Omega$ and using Eqs.~(\ref{9})--(\ref{12}) and (\ref{2}), we obtain the mean energy of the system as
\begin{equation}\label{13}
  \langle H\rangle= \Omega - T \frac{\partial \Omega}{\partial T} + \mu \langle N\rangle = \sum\limits_{i} \delta_{V,V_{i}}  p_{i} E_{i}.
\end{equation}
Taking the first derivative of the function $\Omega$ with respect to the chemical potential $\mu$ and using Eqs.~(\ref{9}), (\ref{10}), and the relations $\partial E_{i}/\partial \mu=0$, $\partial N_{i}/\partial \mu=0$ we obtain the mean number of particles of the system as
\begin{equation}\label{14}
   \langle N\rangle =-\frac{\partial \Omega}{\partial \mu} = \sum\limits_{i} \delta_{V,V_{i}}  p_{i} N_{i}.
\end{equation}

The first derivative of the function $\Omega$ with respect to the volume $V$ gives the pressure
\begin{eqnarray}\label{15}
   p =-\frac{\partial \Omega}{\partial V} &=& -\sum\limits_{i} \delta_{V,V_{i}}  p_{i} \left(\frac{\partial E_{i}}{\partial V}-\mu \frac{\partial N_{i}}{\partial V}\right) \nonumber \\ &+& T \sum\limits_{i} \frac{\partial \delta_{V,V_{i}}}{\partial V} p_{i}^{1+\frac{1}{z}}.
\end{eqnarray}
Note that in many applications the last term in Eq.~(\ref{15}) is zero. Taking the first derivative of the function $\Omega$ with respect to the variable $z$ and using Eqs.~(\ref{9}), (\ref{10}) and the relations $\partial E_{i}/\partial z=0$, $\partial N_{i}/\partial z=0$ we obtain the conjugate force to the variable $z$ as
\begin{equation}\label{16}
   X =-\frac{\partial \Omega}{\partial z} = T\sum\limits_{i} \delta_{V,V_{i}}  p_{i} \left[1-p_{i}^{1/z}(1-\ln p_{i}^{1/z}) \right].
\end{equation}
In comparison with the Boltzmann-Gibbs statistics, the Tsallis statistics contains a nonvanishing additional conjugate force $X$ defined by Eq.~(\ref{16})~\cite{Parv2,Parv2a}.

Equations (\ref{11})--(\ref{16}) and the Legendre transform (\ref{3}) satisfy the differential thermodynamic relation
\begin{equation}\label{17}
    d\Omega = -S dT - p dV- X dz - \langle N\rangle d\mu
\end{equation}
and the fundamental equation of thermodynamics~\cite{Vives,Parv1}
\begin{equation}\label{18}
    T dS= d\langle H\rangle + pdV +X dz -\mu d\langle N\rangle.
\end{equation}

The fundamental equation of thermodynamics (\ref{18}) provides the first and second principles of thermodynamics
\begin{equation}\label{19}
    \delta Q=TdS, \qquad \delta Q= d\langle H\rangle + pdV +X dz -\mu d\langle N\rangle,
\end{equation}
where $\delta Q$ is a heat transfer by the system to the environment during a quasistatic transition of the system from one equilibrium state to a nearby one.
Then the heat capacity can be defined from the general rule, $\delta Q =C dT$, and the first and second laws of thermodynamics (\ref{19}). In the grand canonical ensemble the heat capacity at the fixed values of $(T,V,z,\mu)$ can be written as
\begin{equation}\label{20} \nonumber
    C_{Vz\mu} = \frac{\delta Q}{dT} = T\frac{\partial S}{\partial T}=\frac{\partial \langle H\rangle}{\partial T} -\mu \frac{\partial \langle N\rangle}{\partial T}
    =-T\frac{\partial^{2} \Omega}{\partial T^{2}} .
\end{equation}

In the grand canonical ensemble the statistical formalism is thermodynamically self-consistent if the thermodynamic potential $\Omega$ is a homogeneous function of the first degree with respect to the extensive variables of state $(V,z)$, i.e., if
\begin{equation}\label{21}
    \Omega(T,V,z,\mu) = V \omega(T,z_{v},\mu),
\end{equation}
where $z_{v}=z/V$. In this case from Eqs.~(\ref{15}), (\ref{16}), (\ref{21}) and (\ref{3}) we obtain the relation between the pressure and the thermodynamic potential and the Euler theorem as
\begin{equation}\label{22}
    \omega=-p - Xz_{v}, \qquad Ts =\varepsilon+p+Xz_{v}-\mu \rho,
\end{equation}
where $\omega=\Omega/V$, $s=S/V$, $\varepsilon=\langle H\rangle/V$ and $\rho=\langle N\rangle/V$. Relations (\ref{21}) and (\ref{22}) can be satisfied in the thermodynamic limit, i.e. when $V\to\infty$, $|z|\to\infty$ and $z_{v}=z/V=const$. Next, to verify this we shall study a special case of this theory, the nonrelativistic ideal gas of hadrons in the grand canonical ensemble, in more detail.

\section{Nonrelativistic hadron gas: Grand canonical ensemble}\label{sec:2}

Let us consider the nonrelativistic ideal gas of the Maxwell-Boltzmann particles in the framework of the Tsallis statistics in the grand canonical ensemble $(T,V,z,\mu)$. In this paper, we investigate the ideal gas only for the values of $z<-1$. To obtain the exact results, we use the integral representation of the Gamma-function~\cite{Abramowitz}
\begin{equation}\label{b1}
  x^{-y} = \frac{1}{\Gamma(y)} \int\limits_{0}^{\infty}  t^{y-1} e^{-tx}  dt , \; Re(x)>0, \; Re(y)>0.
\end{equation}
Substituting Eq.~(\ref{b1}) into (\ref{10}) and using the partition function of the Maxwell-Boltzmann nonrelativistic ideal gas in the Boltzmann-Gibbs statistics
\begin{equation}\label{b1a}
  Z=\exp\left[gV\left(\frac{mT}{2\pi}\right)^{3/2}e^{\mu/T}\right],
\end{equation}
we obtain
\begin{eqnarray}\label{b2}
 1 &=& \sum\limits_{N=0}^{N_{0}} \frac{\tilde{\omega}^{N}}{N!} h(0) \left[1+\frac{1}{z+1}\frac{\Lambda+\mu N}{T}\right]^{z+\frac{3}{2}N}, \\ \label{b2a}
   h(\xi) &=& \frac{(-z-1)^{\frac{3}{2}N}\Gamma(-z-\xi-\frac{3}{2}N)}{\Gamma(-z-\xi)},
\end{eqnarray}
where $\tilde{\omega}=gV(mT/2\pi)^{3/2}$, $g$ is the spin-isospin degeneracy factor and $m$ is the particle mass. The upper limit $N_{0}$ in the sum (\ref{b2}) is determined from the conditions $-z-3N/2>0$, $1+(\Lambda+\mu N)/((z+1)T)>0$ or/and the minimum of the function
\begin{equation}\label{b3}
  \phi(N) = \frac{\tilde{\omega}^{N}}{N!} h(0) \left[1+\frac{1}{z+1}\frac{\Lambda+\mu N}{T}\right]^{z+\frac{3}{2}N}.
\end{equation}
The solution of Eq.~(\ref{b2}) is the norm function $\Lambda$ as a function of the variables of state.

Combining Eqs.~(\ref{9}) and (\ref{b1}) we see that the thermodynamic potential (\ref{11}) for the nonrelativistic ideal gas in the grand canonical ensemble becomes
\begin{eqnarray}\label{b4}
  \Omega &=& \Lambda+T \nonumber \\
  &-& T \sum\limits_{N=0}^{N_{0}} \frac{\tilde{\omega}^{N}}{N!} h(1) \left[1+\frac{1}{z+1}\frac{\Lambda+\mu N}{T}\right]^{z+1+\frac{3}{2}N}. \;\;
\end{eqnarray}
Substituting Eq.~(\ref{b4}) into the differential equation (\ref{14}) and using Eqs.~(\ref{b2}), (\ref{b2a}) we have
\begin{equation}\label{b5}
  \langle N \rangle = \sum\limits_{N=0}^{N_{0}} N \frac{\tilde{\omega}^{N}}{N!} h(0)  \left[1+\frac{1}{z+1}\frac{\Lambda+\mu N}{T}\right]^{z+\frac{3}{2}N}.
\end{equation}
Substituting Eq.~(\ref{b4}) into the differential equation (\ref{13}) and using Eqs.~(\ref{b2}), (\ref{b2a}) and (\ref{b5}) we obtain the average energy of the system in the grand canonical ensemble as
\begin{equation}\label{b6}
  \langle H \rangle = \frac{3}{2}T\sum\limits_{N=0}^{N_{0}} N \frac{\tilde{\omega}^{N}}{N!} h(1) \left[1+\frac{1}{z+1}\frac{\Lambda+\mu N}{T}\right]^{z+1+\frac{3}{2}N}. \;\;\;
\end{equation}
Making use of Eqs.~(\ref{15}), (\ref{b2}), (\ref{b2a}) and (\ref{b4}) we can write the pressure of the nonrelativistic ideal gas in the grand canonical ensemble as
\begin{equation}\label{b7}
  p = \frac{T}{V}\sum\limits_{N=0}^{N_{0}} N \frac{\tilde{\omega}^{N}}{N!} h(1) \left[1+\frac{1}{z+1}\frac{\Lambda+\mu N}{T}\right]^{z+1+\frac{3}{2}N}. \;\;\;\;\;
\end{equation}
Comparing Eqs.~(\ref{b6}) and (\ref{b7}) we obtain the equation of state for the nonrelativistic ideal gas in the grand canonical ensemble
\begin{equation}\label{b8}
  p=\frac{2}{3} \frac{\langle H \rangle}{V}.
\end{equation}
Substituting Eq.~(\ref{b4}) into the differential equation (\ref{12}) and using Eqs.~(\ref{b6}) we can write the entropy of the system in the grand canonical ensemble as
\begin{equation}\label{b9}
  S=-\frac{1}{T} (\Omega-\langle H \rangle+\mu \langle N \rangle).
\end{equation}
This equation coincides with the Legendre transform (\ref{3}). Making use of Eqs.~(\ref{16}) and (\ref{b4}) we can write the conjugate force to the variable $z$ in the form 
\begin{equation}\label{b10}
  \frac{X}{T} = 1+\sum\limits_{N=0}^{N_{0}} \frac{\tilde{\omega}^{N}}{N!} h(1)\zeta \left[1+\frac{1}{z+1}\frac{\Lambda+\mu N}{T}\right]^{z+1+\frac{3}{2}N}, \;\;\;\;\;\;\;\;
\end{equation}
where
\begin{eqnarray}\label{b10a}
 \zeta &=& -1+\ln\left[1+\frac{1}{z+1}\frac{\Lambda+\mu N}{T}\right]\nonumber \\
 &-& \psi(-z-1-\frac{3}{2}N) +\psi\left(-z-1\right).
\end{eqnarray}

The mean occupation numbers for the Maxwell-Boltzmann nonrelativistic ideal gas in the grand canonical ensemble in the framework of the Tsallis statistics can be defined as
\begin{eqnarray}\label{b11}
  \langle n_{\vec{p}\sigma}\rangle &=& \sum\limits_{\{n_{\vec{p}\sigma}\}} n_{\vec{p}\sigma} \frac{1}{\prod\limits_{\vec{p}\sigma}n_{\vec{p}\sigma}!} \nonumber \\
  &\times& \left[1+\frac{1}{z+1}\frac{\Lambda-\sum_{\vec{p}\sigma}n_{\vec{p}\sigma}(\varepsilon_{\vec{p}}-\mu)}{T}\right]^{z},
\end{eqnarray}
where $\varepsilon_{\vec{p}}=\vec{p}^{2}/2m$. Substituting Eq.~(\ref{b1}) into (\ref{b11}) and using the partition function (\ref{b1a}) and the mean occupation numbers of the Maxwell-Boltzmann nonrelativistic ideal gas in the Boltzmann-Gibbs statistics
\begin{equation}\label{b12}
  \langle n_{\vec{p}\sigma}\rangle=e^{-\frac{\varepsilon_{\vec{p}}-\mu}{T}},
\end{equation}
we obtain
\begin{eqnarray}\label{b13}
  \langle n_{\vec{p}\sigma}\rangle &=& \sum\limits_{N=0}^{N_{0}} \frac{\tilde{\omega}^{N}}{N!} h(0) \nonumber \\
  &\times& \left[1+\frac{1}{z+1}\frac{\Lambda+\mu N-(\varepsilon_{\vec{p}}-\mu)}{T}\right]^{z+\frac{3}{2}N},
\end{eqnarray}
where $\Lambda$ is calculated from Eqs.~(\ref{b2}) and (\ref{b2a}).

\section{Nonrelativistic hadron gas in the thermodynamic limit: Grand canonical ensemble}\label{sec:3}
Let us rewrite the thermodynamic quantities of the nonrelativistic ideal gas given above in the thermodynamic limit
\begin{equation}\label{c1}
  V\to\infty, \quad |z|\to\infty, \quad z_{v}=\frac{z}{V}=const.
\end{equation}
As we expected, the sum (\ref{b2}) in the thermodynamic limit (\ref{c1}) can be approximated by its maximal term, which gives the main contribution in (\ref{b2}). Thus, from the maximum of the function $\phi(N)$ (\ref{b3}) and Eq.~(\ref{b2}) we have two equations, $\ln \phi(N)=0$ and $\partial\ln \phi(N)/\partial N=0$, which can be rewritten as
\begin{eqnarray}\label{c2}
\frac{5}{2} +\ln\frac{\omega'}{\rho}+ \left(\frac{z_{v}}{\rho}+\frac{3}{2}\right) \ln\frac{T_{eff}}{T}   &=& 0, \\ \label{c3}
  \ln\frac{\omega'}{\rho}+ \frac{3}{2}\ln\frac{T_{eff}}{T}+ \frac{\mu}{T_{eff}} &=& 0
\end{eqnarray}
and
\begin{equation}\label{c3a}
 T_{eff} \equiv T \ \frac{1+\frac{1}{z_{v}} \frac{\Lambda_{v}+\mu \rho}{T}}{1+\frac{3}{2}\frac{\rho}{z_{v}}},
\end{equation}
where $\omega'\equiv\tilde{\omega}/V=g (mT/2\pi)^{3/2}$ and $\Lambda_{v}=\Lambda/V$. Here $\langle N\rangle = N$ corresponds to the maximal term in the sum (\ref{b2}). From Eq.~(\ref{c2}) we obtain the norm function $\Lambda$ as
\begin{eqnarray}\label{c4}
  \Lambda_{v} &=& -\mu \rho - z_{v} T \left[1-\left(1+\frac{3}{2}\frac{\rho}{z_{v}}\right)\frac{T_{eff}}{T} \right], \\ \label{c4a}
  T_{eff} &=& T \left(\tilde{Z}_{G}e^{3/2}\right)^{-\frac{1}{\frac{z_{v}}{\rho}+\frac{3}{2}}},
\end{eqnarray}
where $\tilde{Z}_{G}=\omega'e/\rho=(g e/\rho) (mT/2\pi)^{3/2}$~\cite{Parv2a}. Substituting Eq.~(\ref{c2}) into Eq.~(\ref{c3}) and using Eq.~(\ref{c3a}) we obtain the equation for the hadron density $\rho$ as
\begin{equation}\label{c5}
  \mu = \left[\frac{5}{2}+\frac{z_{v}}{\rho} \ln\frac{T_{eff}}{T}\right] T_{eff}.
\end{equation}
Its solution is the variable $\rho$ as a function of the variables of state $(T,z_{v},\mu)$. Taking Eq.~(\ref{b4}) in the thermodynamic limit (\ref{c1}) and considering Eq.~(\ref{c2}) the density of the thermodynamic potential can be written as
\begin{equation}\label{c6}
  \omega = \Lambda_{v}= -\mu \rho - z_{v} T \left[1-\left(1+\frac{3}{2}\frac{\rho}{z_{v}}\right)\frac{T_{eff}}{T} \right].
\end{equation}
Considering Eqs.~(\ref{13}), (\ref{c5}) and (\ref{c6}) we can write the energy density in the thermodynamic limit as
\begin{equation}\label{c7}
 \varepsilon =-T^{2} \frac{\partial}{\partial T} \left(\frac{\omega}{T}\right) +\mu \rho = \frac{3}{2} \rho  T_{eff}.
\end{equation}
Using Eqs.~(\ref{16}), (\ref{c5}) and (\ref{c6}) we obtain the conjugate force to the variable $z$ in the thermodynamic limit
\begin{equation}\label{c8}
  X = - \frac{\partial \omega}{\partial z_{v}} = T \left\{1 - \frac{T_{eff}}{T} \left[1-\ln\frac{T_{eff}}{T}\right]\right\}.
\end{equation}
Substituting Eqs.~(\ref{21}), (\ref{c6}) into differential equation (\ref{15}) we can write the pressure in the thermodynamic limit as
\begin{equation}\label{c9}
  p=-\omega + z_{v}\frac{\partial \omega}{\partial z_{v}} = \frac{2}{3} \varepsilon =  \rho  T_{eff}.
\end{equation}
Making use of Eqs.~(\ref{b9}), (\ref{c6}) and (\ref{c7}) we can write the entropy density in the thermodynamic limit as
\begin{equation}\label{c10}
  s=-\frac{\omega - \varepsilon + \mu \rho}{T} =z_{v} \left[1-\frac{T_{eff}}{T} \right].
\end{equation}
Using equation $\ln \phi(N)=0$ we obtain the mean occupation numbers (\ref{b13}) in the thermodynamic limit (\ref{c1}) as
\begin{equation}\label{c11}
  \langle n_{\vec{p}\sigma}\rangle = e^{-\frac{\varepsilon_{\vec{p}}-\mu}{T_{eff}}}.
\end{equation}
Then, the one-particle distribution function can be defined as
\begin{equation}\label{c13}
  f(\vec{p})=\frac{1}{(2\pi)^{3}\rho} \sum\limits_{\sigma} \langle n_{\vec{p}\sigma}\rangle = \frac{g}{(2\pi)^{3}\rho} \ e^{-\frac{\varepsilon_{\vec{p}}-\mu}{T_{eff}}},
\end{equation}
where
\begin{equation}\label{c14a}
   \int d^{3}\vec{p} f(\vec{p})=1.
\end{equation}

Let us verify that the nonrelativistic ideal gas in the grand canonical ensemble satisfies the principle of additivity and the zeroth law of thermodynamics in the thermodynamic limit. Suppose that the system is divided into two subsystems ($1$ and $2$). Then, the extensive variables of state of the grand canonical ensemble are additive
\begin{equation}\label{c14}
  V=V^{1}+V^{2}, \qquad z=z^{1}+z^{2}
\end{equation}
and the intensive variables of state of the grand canonical ensemble are the same in each subsystem
\begin{equation}\label{c15}
  T=T^{1}=T^{2}, \quad z_{v}=z_{v}^{1}=z_{v}^{2}, \quad  \mu=\mu^{1}=\mu^{2}.
\end{equation}

For the homogeneous functions of the first order, $\mathcal{F}=(\Lambda,\Omega,\langle H\rangle,\langle N\rangle,S)$, and their densities, $\varphi=(\Lambda_{v},\omega,\varepsilon,\rho,s)$, we have the relation
\begin{equation}\label{c16}
  \mathcal{F}(T,V,z,\mu)=V \varphi(T,z_{v},\mu)
\end{equation}
and the Euler theorem
\begin{equation}\label{c17}
   V\frac{\partial \mathcal{F}}{\partial V} + z\frac{\partial \mathcal{F}}{\partial z} = \mathcal{F}.
\end{equation}
Using Eqs.~(\ref{c4}), (\ref{c4a}), (\ref{c6}), (\ref{c7}), (\ref{c10}) and Eqs.~(\ref{c14}), (\ref{c15}) we can verify that the homogeneous functions of the first order $\mathcal{F}$ are extensive and their densities $\varphi$, which are the homogeneous functions of the zero order, are intensive
\begin{equation}\label{c18}
  \mathcal{F} = \mathcal{F}^{1} +\mathcal{F}^{2}, \qquad \varphi = \varphi^{1} =\varphi^{2}.
\end{equation}

For the homogeneous functions of the zero order, $\phi=(p,X)$, we have the relation
\begin{equation}\label{c19}
  \phi(T,V,z,\mu)=\phi(T,z_{v},\mu)
\end{equation}
and the Euler theorem
\begin{equation}\label{c20}
   V\frac{\partial \phi}{\partial V} + z\frac{\partial \phi}{\partial z} = 0.
\end{equation}
Using Eqs.~(\ref{c5}), (\ref{c8}), (\ref{c9}) and Eqs.~(\ref{c14}), (\ref{c15}) we obtain that the homogeneous functions of the zero order $\phi$ are intensive
\begin{equation}\label{c21}
  \phi = \phi^{1} =\phi^{2}.
\end{equation}
These relations prove the zero law of thermodynamics for the Tsallis statistics in the grand canonical ensemble.

\section{Ensemble equivalence}\label{sec:4}
\subsection{Equivalence of canonical and grand canonical ensembles}
Let us demonstrate that the thermodynamic quantities of the canonical ensemble in the thermodynamic limit can be obtained from the quantities of the grand canonical ensemble by the  transformation of the variables of state $(T,z_{v},\mu)$ with the variables of state of the canonical ensemble $(T,v,\tilde{z})$, where $\tilde{z}\equiv z/\langle N\rangle=z_{v}/\rho$ and $v\equiv V/\langle N\rangle=1/\rho$. Expressing Eqs.~(\ref{c4a}), (\ref{c5}) in terms of the variable $\tilde{z}$, we obtain the chemical potential of the canonical ensemble in the thermodynamic limit (see Eq.~(38) in~\cite{Parv2a})
\begin{eqnarray}\label{e1}
  \mu &=& \left[\frac{5}{2}+\tilde{z} \ln\frac{T_{eff}}{T}\right] T_{eff}, \\ \label{e1a}
  T_{eff} &=& T \left(\tilde{Z}_{G}e^{3/2}\right)^{-\frac{1}{\tilde{z}+\frac{3}{2}}}.
\end{eqnarray}
The energy density (\ref{c7}) of the grand canonical ensemble can be transformed to the energy per particle of the canonical ensemble, $\tilde{\varepsilon}\equiv \langle H\rangle/\langle N\rangle$, by using the variable $\tilde{z}$ (see Eq.~(32) in~\cite{Parv2a})
\begin{equation}\label{e2}
 \tilde{\varepsilon}=\frac{\varepsilon}{\rho} = \frac{3}{2} T_{eff}.
\end{equation}
The thermodynamic potential of the canonical ensemble, the free energy, can be obtained from the thermodynamic potential of the grand canonical ensemble by the Legendre transform, i.e. $F=\Omega+\mu \langle N\rangle$. Then, from Eq.~(\ref{c6}) we have the thermodynamic potential per particle of the canonical ensemble, $\tilde{f}\equiv F/\langle N\rangle$, which can be written as (see Eqs.~(30) and (33) in~\cite{Parv2a})
\begin{equation}\label{e3}
  \tilde{f}= \frac{\omega}{\rho}+\mu = -\tilde{z} T \left[1-\left(1+\frac{3}{2\tilde{z}}\right) \frac{T_{eff}}{T}\right].
\end{equation}

Rewriting the variable $X$ of the grand canonical ensemble (\ref{c8}) in terms of $\tilde{z}$, we obtain the variable $X$ of the canonical ensemble (see Eq.~(37) in~\cite{Parv2a})
\begin{equation}\label{e4}
  X =  T \left\{1 - \frac{T_{eff}}{T} \left[1-\ln\frac{T_{eff}}{T}\right]\right\}.
\end{equation}
The pressure of the grand canonical ensemble (\ref{c9}) can be rewritten as
\begin{equation}\label{e5}
  p= \frac{2}{3}\frac{\tilde{\varepsilon}}{v}  =  \frac{T_{eff}}{v}.
\end{equation}
This equation for the pressure of the canonical ensemble exactly coincides with Eq.~(36) in~\cite{Parv2a}. Expressing Eq.~(\ref{c10}) in terms of the variable $\tilde{z}$ we obtain the entropy per particle of the canonical ensemble, $\tilde{s}\equiv S/\langle N\rangle$, in the form (see Eq.~(34) in~\cite{Parv2a})
\begin{equation}\label{e6}
  \tilde{s}=\frac{s}{\rho} =\tilde{z}\left[1-\frac{T_{eff}}{T} \right].
\end{equation}

Changing the variables of state in Eqs.~(\ref{c11})--(\ref{c13}) and using Eq.~(\ref{e1}) we obtain the mean occupation numbers and the distribution function $f(\vec{p})$ for the canonical ensemble
\begin{equation}\label{e7}
  \langle n_{\vec{p}\sigma}\rangle = \frac{1}{gv} \left(\frac{mT_{eff}}{2\pi}\right)^{-3/2} e^{-\frac{\vec{p}^{2}}{2mT_{eff}}}
\end{equation}
and
\begin{equation}\label{e8}
 f(\vec{p}) = \left(\frac{1}{2\pi m_{eff} T}\right)^{3/2}  \ e^{-\frac{\vec{p}^{2}}{2m_{eff} T}},
\end{equation}
where
\begin{equation}\label{e9}
   \frac{T_{eff}}{T} =\frac{m_{eff}}{m}= \left(\tilde{Z}_{G}e^{3/2}\right)^{-\frac{1}{\tilde{z}+\frac{3}{2}}}.
\end{equation}
Equations (\ref{e8}) and (\ref{e9}) exactly coincide with Eqs.~(43) and (44), respectively, given in~\cite{Parv2a}.

Thus, we have proved that the canonical and grand canonical ensembles of the Tsallis statistics are equivalent in the thermodynamic limit.

\begin{figure*}
\centering
\resizebox{0.9\textwidth}{!}{%
  \includegraphics{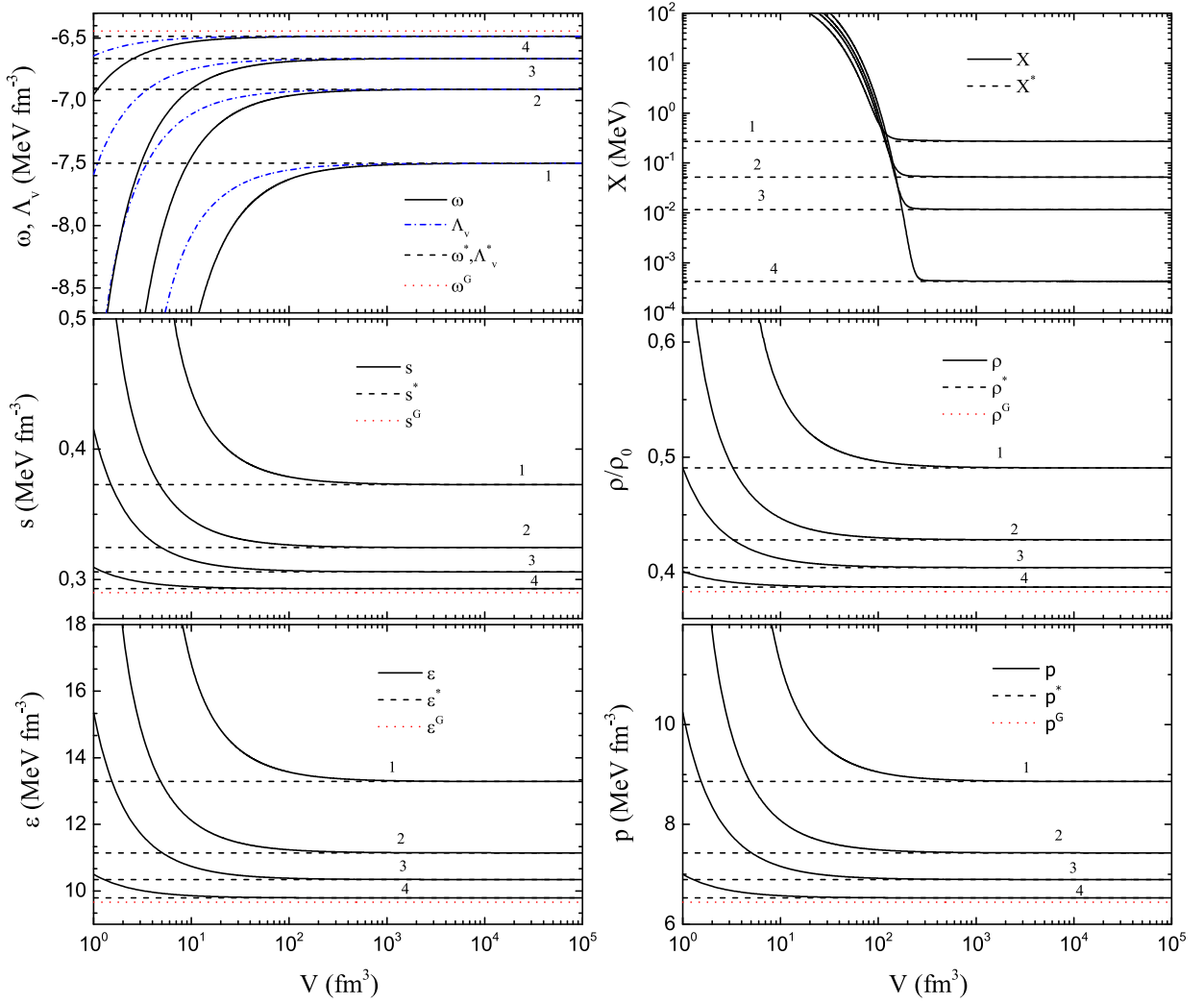}
}
\caption{(Color online) The density of the thermodynamic potential $\omega$, the density of the norm function $\Lambda_{v}$, the entropy density $s$, the energy density $\varepsilon$, the variable $X$, the density $\rho$ and the pressure $p$ as functions of the volume $V$ for the nonrelativistic ideal gas of neutrons in the grand canonical ensemble in the Tsallis statistics at temperature $T=100$ MeV and the chemical potential $\mu=-200$ MeV for different values of $z_{v}$. The curves $1,2,3$ and $4$ correspond to the values of $z_{v}=-5,-10,-20$ and $-100$ fm$^{-3}$, respectively. The dashed lines represent the Tsallis statistics in the thermodynamic limit. The dotted lines represent the Boltzmann-Gibbs statistics and the  dash-dotted lines correspond to the quantity $\Lambda_{v}$. } \label{fig1}
\end{figure*}

\subsection{Equivalence of microcanonical and grand canonical ensembles}
Let us demonstrate that the thermodynamic quantities of the microcanonical ensemble in the thermodynamic limit can be obtained from the quantities of the grand canonical ensemble by the transformation of the variables of state $(T,z_{v},\mu)$ with the variables of state of the microcanonical ensemble $(\tilde{\varepsilon},v,\tilde{z})$, where $\tilde{\varepsilon}\equiv\langle H\rangle/\langle N\rangle=\varepsilon/\rho$. Using the definition of $\tilde{Z}_{G}$ given in Sec.~\ref{sec:3} and Eqs.~(\ref{c7}), (\ref{c4a}), we can write (see Eq.~(48) in~\cite{Parv2})
\begin{equation}\label{e10}
  w=\left(\tilde{Z}_{G}e^{3/2}\right)^{\frac{\tilde{z}}{\tilde{z}+\frac{3}{2}}}=g v \left(\frac{m \tilde{\varepsilon} e^{5/3}}{3\pi}\right)^{3/2}.
\end{equation}
Using again Eqs.~(\ref{c7}) and (\ref{c4a}) we obtain
\begin{eqnarray}\label{e11}
  T &=& \frac{2}{3} \tilde{\varepsilon} w^{1/\tilde{z}}, \\ \label{e12}
  T_{eff} &=& \frac{2}{3} \tilde{\varepsilon}.
\end{eqnarray}
Compare Eq.~(\ref{e11}) with Eq.~(51) given in Ref.~\cite{Parv2}. Substituting Eqs.~(\ref{e11}), (\ref{e12}) into Eq.~(\ref{c10}) we obtain the entropy per particle for the microcanonical ensemble as (see Eq.~(49) in~\cite{Parv2})
\begin{equation}\label{e13}
  \tilde{s}=\frac{s}{\rho} =\tilde{z}\left[1-w^{-1/\tilde{z}}\right].
\end{equation}
Considering Eqs.~(\ref{e11}), (\ref{e12}) and (\ref{c5}) we can write the chemical potential for the microcanonical ensemble as (see Eq.~(54) in Ref.~\cite{Parv2})
\begin{equation}\label{e14}
  \mu = \frac{2}{3} \tilde{\varepsilon} \left[\frac{5}{2}-\ln w\right].
\end{equation}
Substituting Eqs.~(\ref{e11}), (\ref{e12}) into Eq.~(\ref{c8}) we obtain (see Eq.~(55) in Ref.~\cite{Parv2})
\begin{equation}\label{e15}
  X= -\frac{2}{3} \tilde{\varepsilon} \left[1+\ln w^{1/\tilde{z}}-w^{1/\tilde{z}}\right].
\end{equation}
Substituting Eqs.~(\ref{e11}), (\ref{e12}) into Eq.~(\ref{c9}) we obtain the pressure for the microcanonical ensemble as (see Eqs.~(53) and (51) in Ref.~\cite{Parv2})
\begin{equation}\label{e16}
  p=\frac{T}{v} w^{-1/\tilde{z}} = \frac{2}{3} \frac{\tilde{\varepsilon}}{v}.
\end{equation}

Thus, we have proved that the microcanonical and grand canonical ensembles of the Tsallis statistics are equivalent in the thermodynamic limit.

\section{Analysis and results}\label{sec:5}
Let us investigate numerically the thermodynamic limit (\ref{c1}) for the nonrelativistic ideal gas in the grand canonical ensemble in the framework of the Tsallis statistics. We shall increase the volume $V$ at the fixed values of the variables of state $(T,z_{v},\mu)$ and compare the thermodynamic quantities of the ideal gas at constant $V$ with their analytical results in the thermodynamic limit. Note that here and hereafter $\rho_{0}$ is the normal nuclear density and the symbols asterisk and "G" denote the thermodynamic quantities in the thermodynamic limit (\ref{c1}) and in the Boltzmann-Gibbs limit, respectively.

Figure~\ref{fig1} shows the dependence of the density of thermodynamic potential $\omega$, the density of the norm function $\Lambda_{v}$, the entropy density $s$, the energy density $\varepsilon$, the variable $X$, the reduced particle density $\rho/\rho_{0}$ and the pressure $p$ on the volume $V$ for the nonrelativistic ideal gas of neutrons in the grand canonical ensemble in the Tsallis statistics at a given temperature and the chemical potential for different values of the entropic variable $z_{v}$. For the given $T$, $z_{v}$ and $\mu$, the density of the thermodynamic potential $\omega$ (\ref{b4}) and the density of the norm function $\Lambda_{v}$ (\ref{b2}) increase with $V$ and attain the constants, $\omega^{*}\equiv\lim\limits_{V\to\infty,z\to\infty}\omega$, (\ref{c6}) and, $\Lambda_{v}^{*}\equiv\lim\limits_{V\to\infty,z\to\infty}\Lambda_{v}$, (\ref{c4}), respectively, at very large values of $V$. The density of the thermodynamic potential $\omega$ is much lower than the density of the norm function $\Lambda_{v}$ at finite values of $V$ and it tends to $\Lambda_{v}$ with $V$. In the thermodynamic limit as $V\to\infty$ and at $z_{v}$ constant the function $\Lambda_{v}$ achieves the finite value $\omega^{*}$; however, the density of the thermodynamic potential $\omega^{*}$ differs essentially from the density of the thermodynamic potential of the Boltzmann-Gibbs statistics, $\omega^{G}\equiv\lim\limits_{z_{v}\to\infty}\omega^{*}$. At finite values of $z_{v}$ and large values of the volume $V$ the density of the norm function $\Lambda_{v}^{*}$ coincides with the density of the thermodynamic potential $\omega^{*}$ and differs from its Boltzmann-Gibbs limit $\omega^{G}$. The density of the thermodynamic potential $\omega^{*}$ increases with $|z_{v}|$ at fixed variables of state $(T,\mu)$ and attains its Boltzmann-Gibbs value $\omega^{G}$ at very large values of $|z_{v}|\to\infty$. The functions $\omega$ and $\Lambda_{v}$ tend to minus infinity with decreasing the volume $V$.

\begin{figure}
\centering
\resizebox{0.45\textwidth}{!}{%
  \includegraphics{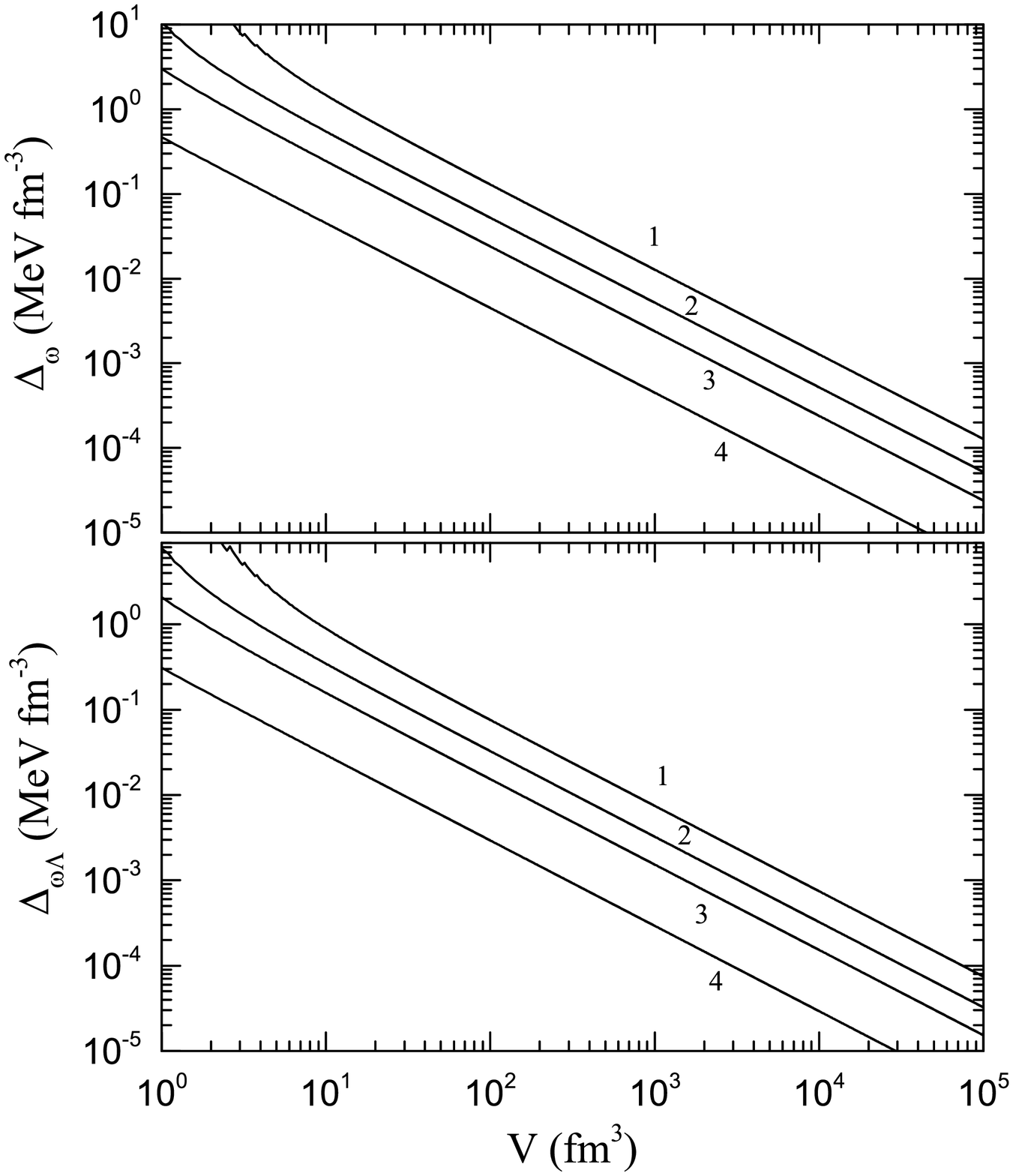}
}
\caption{(Color online) The shift $\Delta_{\omega}=\omega^{*}-\omega$ and the shift $\Delta_{\omega\Lambda}=\Lambda_{v}-\omega$ as functions of the volume $V$ for the nonrelativistic ideal gas of neutrons in the grand canonical ensemble of the Tsallis statistics at the temperature $T=100$ MeV and chemical potential $\mu=-200$ MeV for different values of $z_{v}$. The curves $1,2,3$ and $4$ correspond to the values of $z_{v}=-5,-10,-20$ and $-100$ fm$^{-3}$, respectively.} \label{fig2}
\end{figure}
\begin{figure}
\centering
\resizebox{0.45\textwidth}{!}{%
  \includegraphics{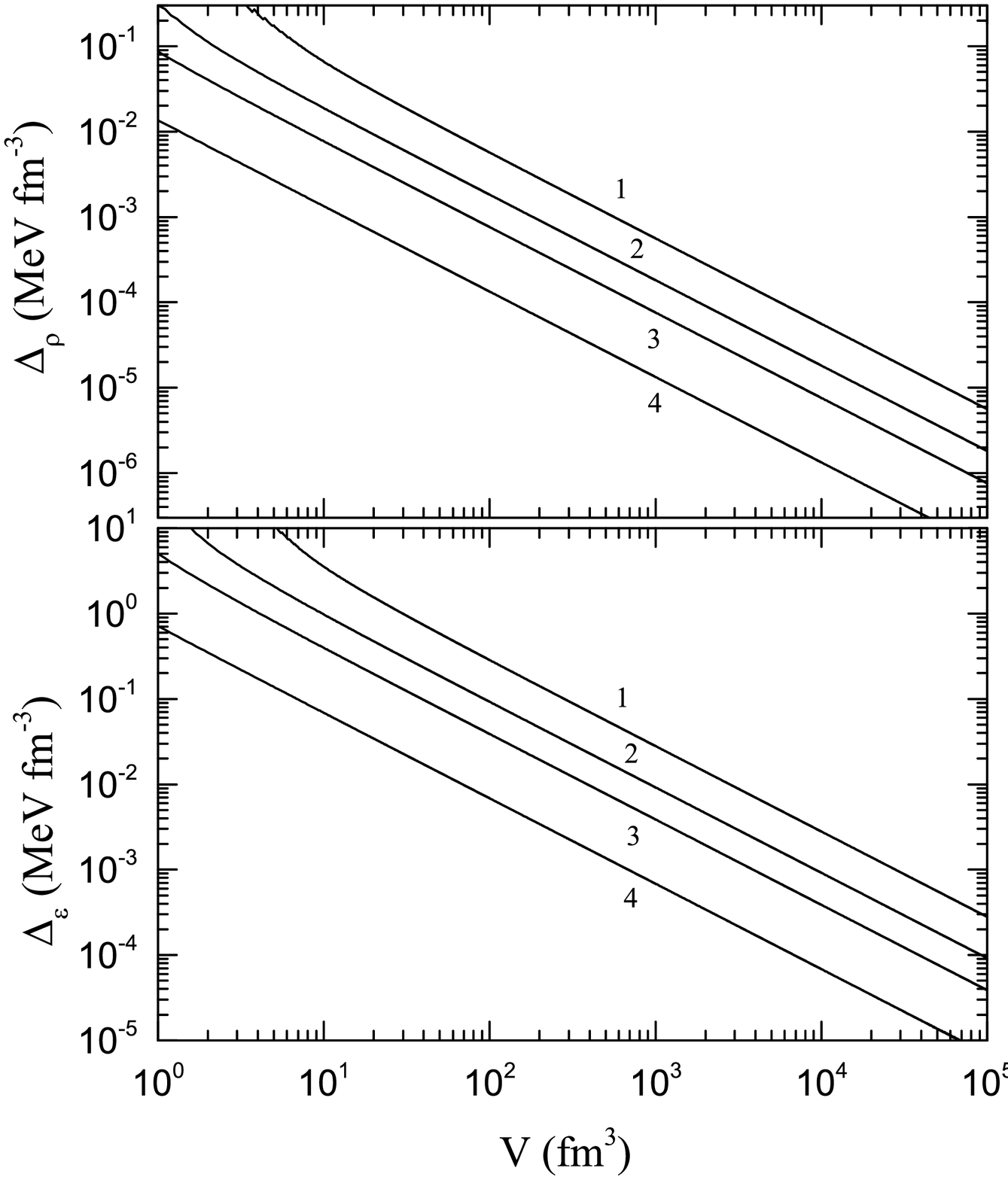}
}
\caption{(Color online) The shift $\Delta_{\rho}=\rho-\rho^{*}$ and the shift $\Delta_{\varepsilon}=\varepsilon-\varepsilon^{*}$ as functions of the volume $V$ for the nonrelativistic ideal gas of neutrons in the grand canonical ensemble of the Tsallis statistics at the temperature $T=100$ MeV and chemical potential $\mu=-200$ MeV for different values of $z_{v}$. The curves $1,2,3$ and $4$ correspond to the values of $z_{v}=-5,-10,-20$ and $-100$ fm$^{-3}$, respectively.} \label{fig3}
\end{figure}

The shape of the entropy density in the thermodynamic limit is shown in the middle left panel of Fig.~\ref{fig1}. The entropy density $s$ (\ref{b9}) decreases with $V$ at fixed values of $(T,z_{v},\mu)$ and attains the constant, $s^{*}\equiv\lim\limits_{V\to\infty,z\to\infty} s$, (\ref{c10}) as $V\to\infty$. At finite values of $z_{v}$, the entropy density in the thermodynamic limit  $s^{*}$ differs essentially from the entropy density of the Boltzmann-Gibbs statistics $s^{G}\equiv\lim\limits_{z_{v}\to\infty}s^{*}$. The entropy density $s^{*}$ decreases with $|z_{v}|$ and achieves the Boltzmann-Gibbs limit only as $|z_{v}|\to\infty$.

The behaviour of the energy density in the thermodynamic limit is shown in the bottom left panel of Fig.~\ref{fig1}. The energy density $\varepsilon$ (\ref{b6}) decreases with $V$ at fixed values of the variables of state $(T,z_{v},\mu)$ and attains the constant value, $\varepsilon^{*}\equiv\lim\limits_{V\to\infty,z\to\infty} \varepsilon$, (\ref{c7}) at $V\to\infty$. At finite values of $z_{v}$, the energy density in the thermodynamic limit  $\varepsilon^{*}$ differs from the energy density of the Boltzmann-Gibbs statistics $\varepsilon^{G}\equiv\lim\limits_{z_{v}\to\infty}\varepsilon^{*}$. The energy density $\varepsilon^{*}$ decreases with $|z_{v}|$ and achieves the Boltzmann-Gibbs limit as $|z_{v}|\to\infty$.

\begin{figure}
\centering
\resizebox{0.4\textwidth}{!}{%
  \includegraphics{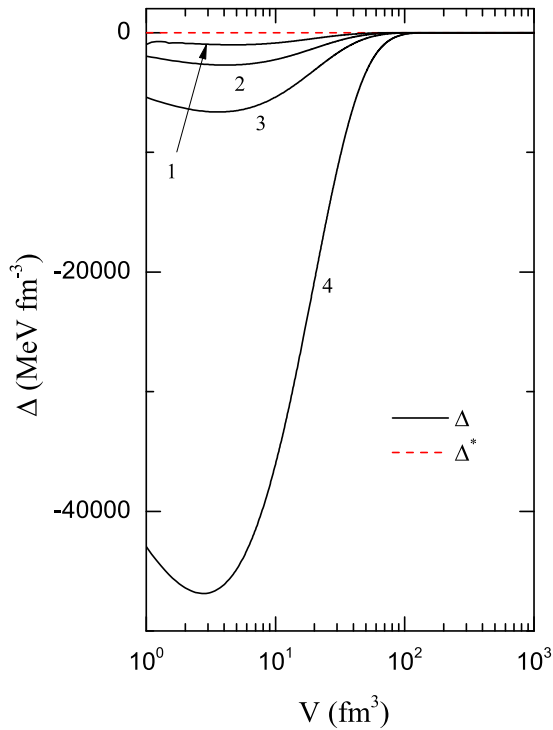}
}
\caption{(Color online) The shift $\Delta=\omega+p+X z_{v}$ as a function of the volume $V$ for the nonrelativistic ideal gas of neutrons in the grand canonical ensemble of the Tsallis statistics at the temperature $T=100$ MeV and chemical potential $\mu=-200$ MeV for different values of $z_{v}$. The curves $1,2,3$ and $4$ correspond to the values of $z_{v}=-5,-10,-20$ and $-100$ fm$^{-3}$, respectively. The dashed line represents the shift $\Delta^{*}=\omega^{*}+p^{*}+X^{*} z_{v}$ in the thermodynamic limit.} \label{fig4}
\end{figure}
\begin{figure}
\centering
\resizebox{0.45\textwidth}{!}{%
  \includegraphics{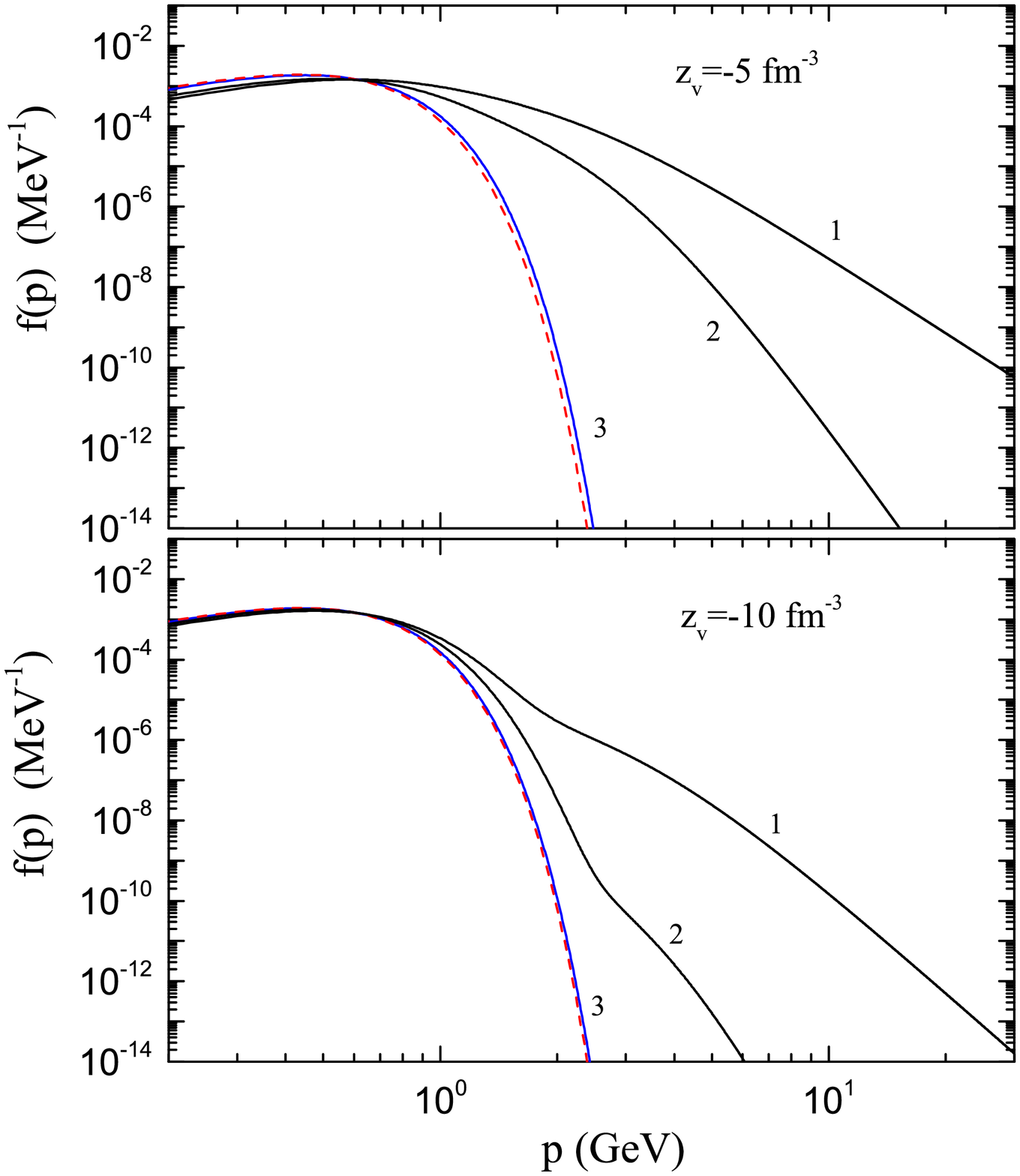}
}
\caption{(Color online) The dependence of the one-particle distribution function on the momentum $p$ for the nonrelativistic ideal gas of neutrons in the grand canonical ensemble of the Tsallis statistics at the temperature $T=100$ MeV and chemical potential $\mu=-200$ MeV for two values of $z_{v}$ and different values of $V$ and $z$. The solid curves $1,2$ and $3$ correspond to the values of $V=1.728,3.375$ fm$^{3}$ and $V=\infty$, respectively. The dashed lines represent the Maxwell-Boltzmann distribution function of the Boltzmann-Gibbs statistics. } \label{fig5}
\end{figure}

The variable $X$ (\ref{b10}) as a function of the volume $V$ at fixed values of the variables of state $(T,z_{v},\mu)$ is shown in the upper right panel of Fig.~\ref{fig1}. It also decreases with $V$ and attains the constant value, $X^{*}\equiv\lim\limits_{V\to\infty,z\to\infty} X$, (\ref{c8}) at $V\to\infty$. For the finite values of $z_{v}$, the function $X^{*}$ is nonvanishing and differs from its Boltzmann-Gibbs zero value, $X^{G}=0$. The conjugate force $X^{*}$ decreases with $|z_{v}|$ and vanishes as $|z_{v}|\to\infty$.

The density of neutrons (\ref{b5}) is shown in the middle right panel of Fig.~\ref{fig1}. It decreases with $V$ and attains the constant value $\rho^{*}$ (\ref{c5}) as $V\to\infty$. For finite values of $z_{v}$, the neutron density $\rho^{*}$ differs from its Boltzmann-Gibbs limit $\rho^{G}$. The density $\rho^{*}$ decreases with $|z_{v}|$ and achieves the Boltzmann-Gibbs limit $\rho^{G}$ as $|z_{v}|\to\infty$.

The shape of the pressure (\ref{b7}) is shown in the bottom right panel of Fig.~\ref{fig1}. It decreases with $V$ and attains the constant value $p^{*}$ (\ref{c9}) as $V\to\infty$. For finite values of $z_{v}$, the pressure $p^{*}$ differs from its Boltzmann-Gibbs limit $p^{G}$. The pressure $p^{*}$ decreases with $|z_{v}|$ and achieves the Boltzmann-Gibbs limit $p^{G}$ as $|z_{v}|\to\infty$.

Figure~\ref{fig2} illustrates the dependence of the shift $\Delta_{\omega}\equiv\omega^{*}-\omega$ and $\Delta_{\omega\Lambda}\equiv\Lambda_{v}-\omega$ on the volume $V$ for the nonrelativistic ideal gas of neutrons in the grand canonical ensemble in the Tsallis statistics at the given temperature and chemical potential for different values of the entropic variable $z_{v}$. At the fixed values of the variables of state $(T,z_{v},\mu)$, the deviation of the density of the thermodynamic potential $\omega$ from its thermodynamic limit value $\omega^{*}$ decreases with $V$ and it vanishes as $V\to\infty$. The behaviour of the shift $\Delta_{\omega}$ with $V$ confirms numerically the thermodynamic limit (\ref{c1}) for the Tsallis statistics in the grand canonical ensemble. The shift $\Delta_{\omega\Lambda}$ also decreases with $V$ and vanishes as $V\to\infty$. Such behavior proves numerically the equality of the density of the thermodynamic potential $\omega^{*}$ with the density of the norm function $\Lambda_{v}^{*}$ in the thermodynamic limit.

Figure~\ref{fig3} shows the dependence of the shift $\Delta_{\rho}\equiv\rho-\rho^{*}$ and the shift $\Delta_{\varepsilon}\equiv\varepsilon-\varepsilon^{*}$ on the volume $V$ for the nonrelativistic ideal gas of neutrons in the grand canonical ensemble of the Tsallis statistics at the given temperature $T$ and chemical potential $\mu$ for different values of $z_{v}$. The functions $\Delta_{\rho}$ and $\Delta_{\varepsilon}$ decrease with $V$ and tend to zero as $V\to\infty$. Such behaviour of the shifts $\Delta_{\rho}$ and $\Delta_{\varepsilon}$ proves numerically the thermodynamic limit (\ref{c1}) for the density $\rho$ and the energy density $\varepsilon$ of the Tsallis statistics.

Let us verify numerically the homogeneity properties of the thermodynamic potential of the nonrelativistic ideal gas of neutrons in the grand canonical ensemble for the Tsallis statistics. We shall find the range of the variables of state $V$ and $z$ at the fixed values of $(T,z_{v},\mu)$ where the thermodynamic quantities of the Tsallis statistics satisfy the main requests of the equilibrium thermodynamics. The formalism of the statistical mechanics in the grand canonical ensemble agrees with the requirements of the equilibrium thermodynamics, i.e. it is thermodynamically self-consistent, if the thermodynamic potential of the grand canonical ensemble is a homogeneous function of the first order with respect to the extensive variables of state. Thus, the Tsallis statistics in the grand canonical ensemble will be thermodynamically self-consistent if its thermodynamic potential is a homogeneous function of the first order with respect to the extensive variables of state $V$ and $z$. This means that the density of the thermodynamic potential $\omega$ for the grand canonical ensemble shall satisfy the following equation:
\begin{equation}\label{x1}
  \Delta=\omega+p+X z_{v}=0
\end{equation}
for any values of the variables of state $(T,z_{v},\mu)$.  This equation is a result of Eqs.~(\ref{21}) and (\ref{22}).

Let us verify numerically Eq.~(\ref{x1}) for the ideal gas of the Tsallis statistics in a finite volume. Figure~\ref{fig4} displays the dependence of the shift $\Delta$ on the volume $V$ for the nonrelativistic ideal gas of neutrons in the grand canonical ensemble of the Tsallis statistics at the given temperature $T$ and chemical potential $\mu$ for different values of $z_{v}$. It is clearly seen that at small values of the volume $V$ the shift $\Delta$ is not equal to zero and the density of the thermodynamic potential $\omega$ is an inhomogeneous function. At large values of volume $V$ in the thermodynamic limit the shift $\Delta$ vanishes and the density of the thermodynamic potential $\omega$ becomes a homogeneous function of the first order with respect to the extensive variables of state $V$ and $z$. Thus, we have demonstrated numerically that the Tsallis statistics in the grand canonical ensemble becomes thermodynamically self-consistent in the thermodynamic limit (\ref{c1}).

Figure~\ref{fig5} represents the dependence of the one-particle distribution function (\ref{b13}) on the momentum $p$ of a particle for the nonrelativistic ideal gas of neutrons in the grand canonical ensemble of the Tsallis statistics at the given temperature $T$, chemical potential $\mu$ and variable $z_{v}$ for different values of $V$ and $z$. At small values of the volume $V$ and the variable $|z_{v}|$, where the Tsallis statistics is thermodynamically inconsistent, the one-particle distribution function of the Tsallis statistics has a power-law form and its shape deviates essentially from the usual Maxwell-Boltzmann distribution function. The Tsallis one-particle distribution function $f(p)$ tends to the distribution function $f^{*}(p)$ in the thermodynamic limit as $V\to\infty$. The Tsallis distribution function in the thermodynamic limit $f^{*}(p)$ has an exponential form and it is close to the Maxwell-Boltzmann distribution function of the Boltzmann-Gibbs statistics, but they do not coincide. The difference between the distribution functions $f(p)$ and $f^{*}(p)$ diminishes with $|z_{v}|$. For small values of the variable $|z_{v}|$, the Tsallis one-particle distribution function in the thermodynamic limit $f^{*}(p)$ can differ essentially  from the Maxwell-Boltzmann distribution function $f^{G}(p)$ of the Boltzmann-Gibbs statistics.

\begin{figure*}
\centering
\resizebox{0.9\textwidth}{!}{%
  \includegraphics{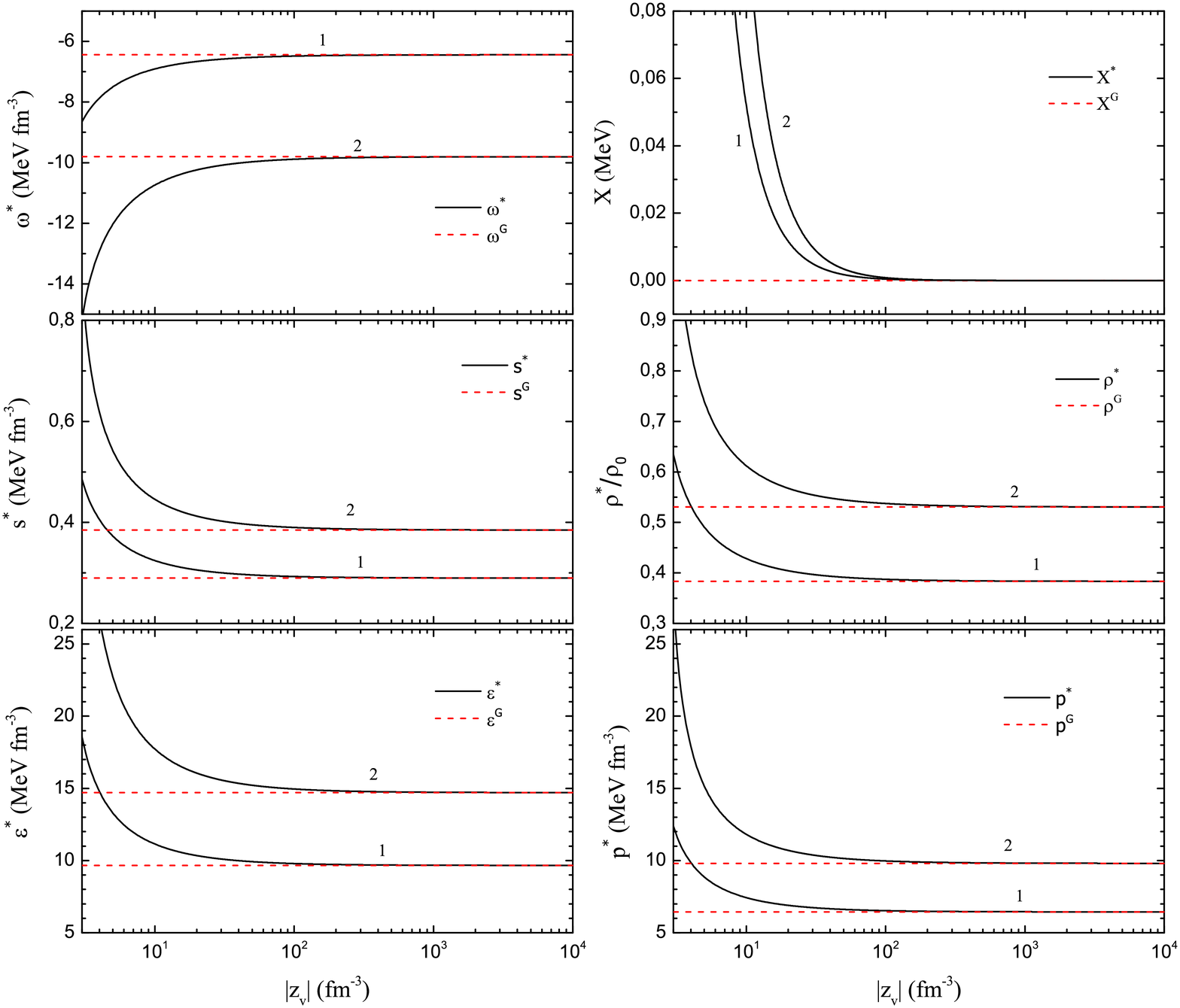}
}
\caption{(Color online) The Boltzmann-Gibbs limit for the Tsallis statistics in the thermodynamic limit. The density of the thermodynamic potential $\omega^{*}$, the entropy density $s^{*}$, the energy density $\varepsilon^{*}$, the variable $X^{*}$, the density $\rho^{*}$ and the pressure $p^{*}$ for the Tsallis statistics in the thermodynamic limit as functions of the variable $z_{v}$ for the nonrelativistic ideal gas of neutrons in the grand canonical ensemble at the chemical potential $\mu=-200$ MeV for different values of temperature $T$. The curves $1$ and $2$ correspond to the values of $T=100$ and $110$ MeV, respectively. The dashed  lines represent the Boltzmann-Gibbs statistics. } \label{fig6}
\end{figure*}

Figure~\ref{fig6} shows the dependence of the density of the thermodynamic potential $\omega^{*}$, the entropy density $s^{*}$, the energy density $\varepsilon^{*}$, the variable $X^{*}$, the density $\rho^{*}$ and the pressure $p^{*}$ for the Tsallis statistics in the thermodynamic limit on the variable $|z_{v}|$ for the nonrelativistic ideal gas of neutrons in the grand canonical ensemble at the given chemical potential $\mu$ for different values of temperature $T$. At small values of $|z_{v}|$, the thermodynamic quantities $\omega^{*}$, $s^{*}$, $\varepsilon^{*}$, $X^{*}$, $\rho^{*}$ and $p^{*}$ of the Tsallis statistics in the thermodynamic limit differ essentially from their values in the Boltzmann-Gibbs limit. Otherwise, at large values of $|z_{v}|$, as $|z_{v}|\to\infty$, these quantities resemble the corresponding thermodynamic quantities of the Boltzmann-Gibbs statistics. The entropy density, the energy density, the neutron density and the pressure for both the Tsallis statistics in the thermodynamic limit and the Boltzmann-Gibbs statistics increase with temperature $T$. However, the density of the thermodynamic potential for both the Tsallis statistics in the thermodynamic limit and the Boltzmann-Gibbs statistics decreases with $T$.

\begin{figure*}
\centering
\resizebox{0.9\textwidth}{!}{%
  \includegraphics{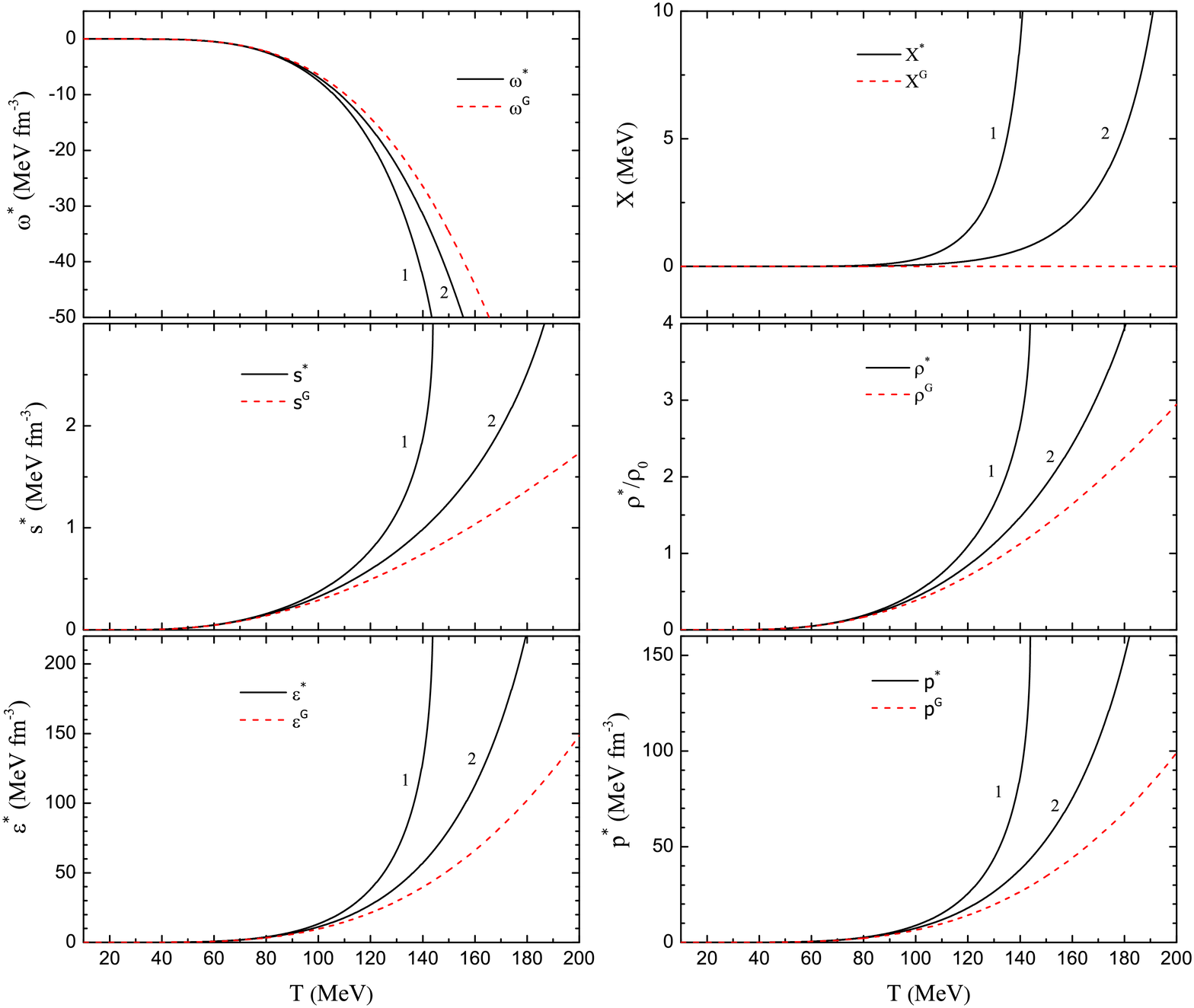}
}
\caption{(Color online) Comparison of the Boltzmann-Gibbs statistics with the Tsallis statistics in the thermodynamic limit. The density of the thermodynamic potential $\omega^{*}$, the entropy density $s^{*}$, the energy density $\varepsilon^{*}$, the variable $X^{*}$, the density $\rho^{*}$ and the pressure $p^{*}$ for the Tsallis statistics in the thermodynamic limit as functions of the temperature $T$ for the nonrelativistic ideal gas of neutrons in the grand canonical ensemble at the chemical potential $\mu=-200$ MeV for different values of the variable $z_{v}$. The solid curves $1$ and $2$ correspond to the values of $z_{v}=-5$ and $-10$ fm$^{-3}$, respectively. The dashed curves represent the Boltzmann-Gibbs statistics.} \label{fig7}
\end{figure*}

Figure~\ref{fig7} represents the dependence of the density of the thermodynamic potential $\omega^{*}$, the entropy density $s^{*}$, the energy density $\varepsilon^{*}$, the variable $X^{*}$, the density $\rho^{*}$ and the pressure $p^{*}$ for the Tsallis statistics in the thermodynamic limit on the temperature $T$ for the nonrelativistic ideal gas of neutrons in the grand canonical ensemble at the given chemical potential $\mu$ for different values of the variable $z_{v}$. At small values of the temperature $T$ and the fixed values of $|z_{v}|$, the Tsallis statistics in the thermodynamic limit approaches the Boltzmann-Gibbs statistics. However, at large values of the temperature $T$, the thermodynamic quantities of the Tsallis statistics in the thermodynamic limit differ from their values in the Boltzmann-Gibbs limit and this deviation increases with $T$. For the negative values of the variable $z_{v}$, the thermodynamic quantities $s^{*}$, $\varepsilon^{*}$, $X^{*}$, $\rho^{*}$ and $p^{*}$ of the Tsallis statistics in the thermodynamic limit overestimate the corresponding quantities of the Boltzmann-Gibbs statistics; however, the density of the thermodynamic potential underestimates the density of the thermodynamic potential of the Boltzmann-Gibbs statistics.

\begin{figure*}
\centering
\resizebox{0.9\textwidth}{!}{%
  \includegraphics{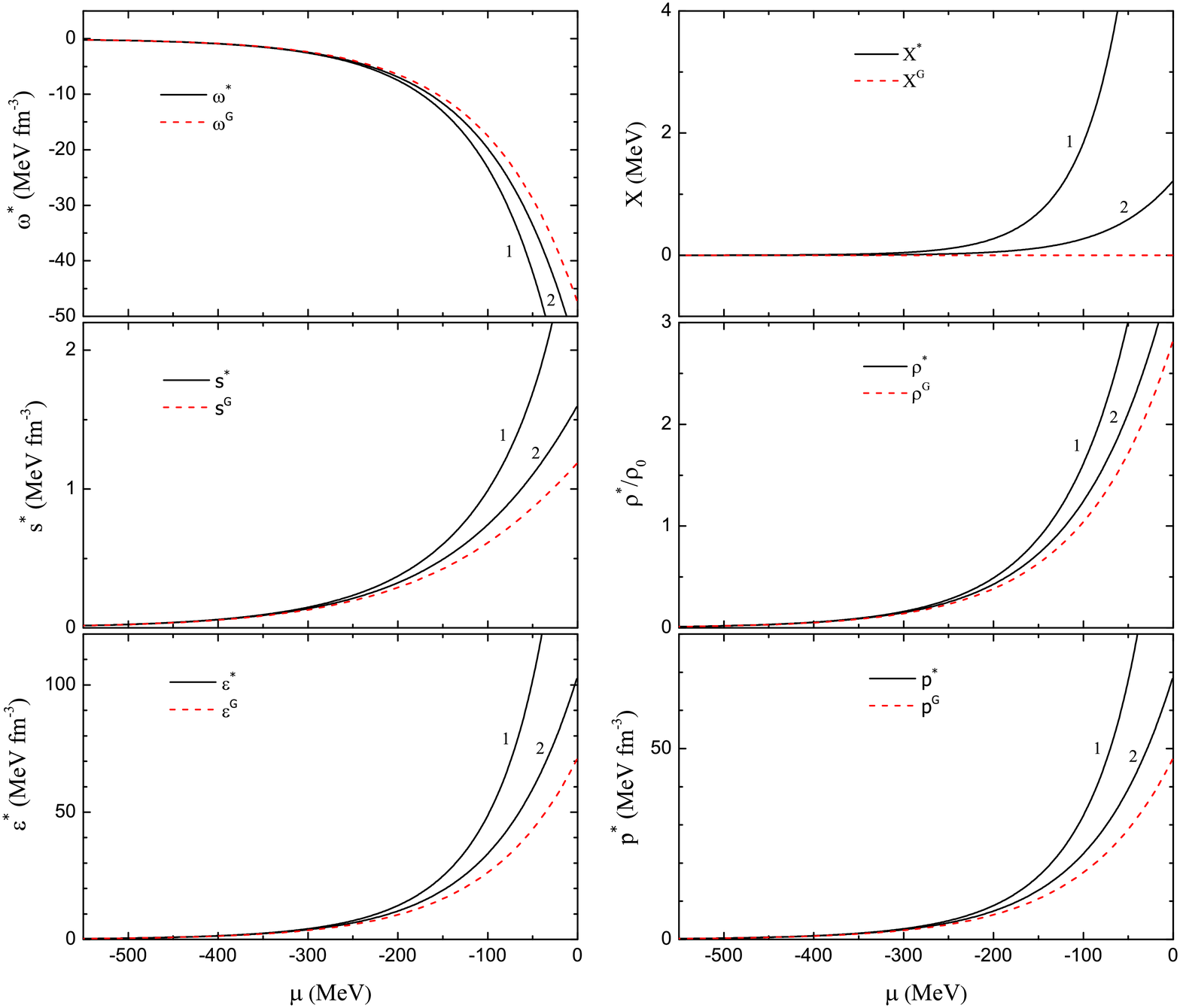}
}
\caption{(Color online) Comparison of the Boltzmann-Gibbs statistics with the Tsallis statistics in the thermodynamic limit. The density of the thermodynamic potential $\omega^{*}$, the entropy density $s^{*}$, the energy density $\varepsilon^{*}$, the variable $X^{*}$, the density $\rho^{*}$ and the pressure $p^{*}$ for the Tsallis statistics in the thermodynamic limit as functions of the chemical potential $\mu$ for the nonrelativistic ideal gas of neutrons in the grand canonical ensemble at the temperature $T=100$ MeV for different values of the variable $z_{v}$. The solid curves $1$ and $2$ correspond to the values of $z_{v}=-5$ and $-10$ fm$^{-3}$, respectively. The dashed curves represent the Boltzmann-Gibbs statistics.} \label{fig8}
\end{figure*}

Figure~\ref{fig8} shows the dependence of the density of the thermodynamic potential $\omega^{*}$, the entropy density $s^{*}$, the energy density $\varepsilon^{*}$, the variable $X^{*}$, the density $\rho^{*}$ and the pressure $p^{*}$ for the Tsallis statistics in the thermodynamic limit on the chemical potential $\mu$ for the nonrelativistic ideal gas of neutrons in the grand canonical ensemble at the given temperature $T$ for different values of the variable $z_{v}$. For the fixed values of $|z_{v}|$, the Tsallis statistics in the thermodynamic limit approaches the Boltzmann-Gibbs statistics at large absolute values of the chemical potential $\mu$. However, at small absolute values of the chemical potential $\mu$, the thermodynamic quantities of the Tsallis statistics in the thermodynamic limit differ from their values in the Boltzmann-Gibbs limit and this deviation increases with $|\mu|\to 0$. For the negative values of the variable $z_{v}$, the thermodynamic quantities $s^{*}$, $\varepsilon^{*}$, $X^{*}$, $\rho^{*}$ and $p^{*}$ of the Tsallis statistics in the thermodynamic limit overestimate the corresponding quantities of the Boltzmann-Gibbs statistics; however, the density of the thermodynamic potential underestimates the density of the thermodynamic potential of the Boltzmann-Gibbs statistics.

\section{Conclusions}\label{sec:6}
In the present paper, the Tsallis statistics in the grand canonical ensemble was formulated in the general form. The thermodynamic potential of the grand canonical ensemble was found from the fundamental thermodynamic potential by the Legendre transform. The probabilities of microstates of the system were derived from the constrained local extrema of the statistical thermodynamic potential of the grand canonical ensemble by the method of the Lagrange multipliers. It was proved that both intensive and extensive thermodynamic quantities of the Tsallis statistics satisfy the same differential thermodynamic relations and statistical definitions in terms of probabilities $p_{i}$ as those for the Boltzmann-Gibbs statistics. This is a consequence of the thermal equilibrium and the Legendre transform. It was shown that in the Tsallis statistics the extensive variable of state $z$ generates the nonvanishing conjugate force $X$. Moreover, the fundamental equation of thermodynamics was consistently derived from the thermodynamic potential of the grand canonical ensemble.

The exact analytical relations for the thermodynamic quantities (thermodynamic potential and its first derivatives) of the nonrelativistic ideal gas in the framework of the Tsallis statistics in the grand canonical ensemble were obtained. They were analytically rewritten in the thermodynamic limit considering the variable $z$ as an extensive variable of state. Based on the example of the nonrelativistic ideal gas in the grand canonical ensemble it was proved that the Tsallis statistics satisfies the requirements of the equilibrium thermodynamics in the thermodynamic limit when the entropic variable $z$ is an extensive variable of state of the system, i.e., the thermodynamic potential is a homogeneous function of the first order with respect to the extensive variables of state of the system. The equivalence of the canonical, grand canonical and microcanonical ensembles in the thermodynamic limit was analytically proved. It was shown that in the thermodynamic limit the relations of the ideal gas of the Tsallis statistics in terms of the function $T_{eff}$ can be written in a form similar to the Boltzmann-Gibbs statistics. The behaviour of the thermodynamic quantities of the ideal gas in both the thermodynamic limit and the Boltzmann-Gibbs limit were also studied. It was numerically proved that the thermodynamic potential of the grand canonical ensemble becomes a homogeneous function of the first order with respect to the extensive variables of state only in the thermodynamic limit. It was shown that the one-particle distribution function has a power-law form when the Tsallis statistics is thermodynamically inconsistent and takes an exponential form in the thermodynamic limit when the Tsallis statistics is thermodynamically consistent.

\begin{acknowledgement}
This work was supported in part by the joint research project of JINR and IFIN-HH, protocol N~4342. I also thank D.-V.~Anghel for helpful comments and discussions.
\end{acknowledgement}

\end{document}